\documentclass[a4paper]{spie}  

\usepackage{amsmath,amsfonts,amssymb}
\usepackage{subcaption}
\usepackage{graphicx}
\usepackage{subcaption}
\usepackage[colorlinks=true, allcolors=blue]{hyperref}
\usepackage{siunitx}
\usepackage{nicefrac}
\usepackage{acronym}

\usepackage{xcolor}
\usepackage{marginnote}

\DeclareSIUnit\clight{\text{\ensuremath{c}}}

\acrodef{PSD}{power spectral density}
\acrodef{FET}{field-effect transistor}
\acrodef{FDA}{floating diffusion amplifier}
\acrodef{FGA}{floating gate amplifier}
\acrodef{MAS}{multi-amplifier sensing}
\acrodef{ROI}{region of interest}
\acrodef{CDS}{correlated double sampling}
\acrodef{DUT}{device under test}
\acrodef{CIC}{clock-induced charges}
\acrodef{CTI}{charge transfer in-efficiency}

\acrodef{LTA}{Low Threshold Acquisition}

\acrodef{DESI}{the Dark Energy Spectroscopic Instrument}
\acrodef{DECam}{the Dark Energy Camera}
\acrodef{LSSTCam}{the Legacy Survey of Space and Time Camera}

\title{Skipper CCD readout time optimization for astronomical applications}

\author[a,*]{Johannes Wüthrich}
\author[a]{Andrin Fazan}
\author[a]{Sean MacBride}
\author[a]{Marcelle Soares-Santos}
\affil[a]{Physik-Institut, Universität Zürich, Winterthurerstrasse 190, 8057 Zürich, Switzerland}

\authorinfo{Further author information: (Send correspondence to J.W.)\\
    \textsuperscript{*}J.W.: E-mail: johannes.wuethrich@physik.uzh.ch}

\begin{document}
\maketitle

\begin{abstract}
Skipper CCDs enable the reduction of CCD readout noise by non-destructively measuring the individual pixel charge packets multiple times.
This readout noise reduction has attracted considerable interest in the astronomical community, particularly in spectroscopic surveys targeting faint objects at high redshifts.
However, noise reduction via repetitive sampling leads to an unavoidable increase in readout time, often to prohibitive levels.
To enable their use in astronomical applications, the optimal operation regime of Skipper CCDs must be determined, balancing noise improvement and readout time.
Traditionally, such optimization has been carried out empirically for each CCD architecture.
We present a general optimization scheme derived from first principles and experimentally verified in the laboratory using a Skipper CCD as used by the Oscura experiment.
While the existence of an optimal combination of correlated double-sampling integration time and number of Skipper samples for reaching a given readout noise level at minimal readout time has previously been observed empirically, we model this trade-off analytically based on the intrinsic noise power spectral density of the sensor, allowing the optimal operating point to be predicted rather than determined experimentally for each architecture.
We further show that the location of this optimum is governed by the per-sample charge-transfer time, whose minimization is therefore key to achieving fast Skipper CCD readout.
\end{abstract}

\keywords{Skipper CCD, Astronomy, Readout Noise, Readout Time}

\section{Introduction}
\label{sec:intro}
Scientific CCDs operated at cryogenic temperatures are the prime sensors used for visible and near infrared light detection on cutting edge ground based astronomical survey telescopes, owing to their high performance and low noise characteristics under these operating conditions.
State-of-the-art astronomical instruments that utilize CCDs include \ac{DESI}~\cite{bebek_status_2017}, \ac{DECam}~\cite{flaugher_dark_2015}, and \ac{LSSTCam}~\cite{kahn_design_2010}.
Fabrication, qualification testing, and operation at low temperatures guarantees negligible dark current in scientific CCDs, and other noise sources such as \ac{CIC} are usually minimized via operational optimizations.
This makes the intrinsic noise from the front-end amplifier, in CCDs usually implemented as \ac{FDA}, the dominating sensor noise source~\cite{janesick_scientific_2001,fernandez_moroni_sub-electron_2012}.
In a Skipper CCD the \ac{FDA} is replaced by a \ac{FGA} and an additional gate is added in the serial register, enabling the charge packets to be moved in and out of the sense node non-destructively~\cite{fernandez_moroni_sub-electron_2012}.
While an \ac{FGA} typically introduces an increased noise \ac{PSD} compared to an \ac{FDA}, and thus a higher intrinsic readout noise, the averaging of $N_{Skip}$ repeated, non-destructive samples of the same charge packets enables the reduction of the readout noise as $\sigma_{RO} \propto \nicefrac{1}{\sqrt{N_{Skip}}}$~\cite{fernandez_moroni_sub-electron_2012}, enabling readout noise levels as low as \SI{0.068}{e^{-}}~\cite{tiffenberg_single-electron_2017}.
This effectively makes Skipper CCDs single photon and single electron sensitive when $N_{Skip}$ is high enough.
The reason for the reduction in readout noise is further discussed in section~\ref{sec:model}.
Given the generally sequential nature of CCD readout, the reduction in readout noise in Skipper CCDs leads to an increased readout time as in first order $t_{RO} \propto N_{Skip}$.
In fundamental science the use of Skipper CCDs is well established in direct detection light dark matter searches, where CCDs are used as an active test mass.
Experiments include DAMIC-M~\cite{settimo_search_2020}, SENSEI~\cite{sensei_collaboration_sensei_2019} and the planned Oscura~\cite{cervantes-vergara_skipper-ccd_2023}.
Given the low mass of CCDs the expected interaction rate with dark matter in these experiments is low, but the use of Skipper CCDs enables very low energy thresholds, accessing a previously unexplored sub-\si{\giga\eV} parameter space for dark matter~\cite{essig_progress_2024}.
For these experiments the long Skipper CCD readout times are acceptable, given the long exposure times, on the order of multiple hours~\cite{cervantes-vergara_skipper-ccd_2023}.

A reduction of readout noise is also beneficial in the context of astronomical applications (spectroscopic or photometric), especially for observations of faint sources, where the sensor readout noise dominates over other noise terms (such as atmospheric noise and shot noise), including for Lyman-$\alpha$ measurements at short wavelengths~\cite{tiffenberg_single-electron_2017,drlica-wagner_characterization_2020}.
Unlike direct dark matter experiments, the increase in readout time in Skipper CCDs can not be ignored for astronomical applications.
Telescope observations are fundamentally limited by the allocated observation time, and any increase in readout time will lead to a decrease in exposure time, as exposure and readout cannot be parallelized when using scientific CCDs.
Thus the optimization of the sensor readout time is the fundamental trade-off vs. the reduction of the readout noise when using Skipper CCDs in astronomical applications.
Previous work considers the use of \ac{ROI} readout modes, where only the parts of the sensor which contain the faint sources are read out using skipper sampling~\cite{drlica-wagner_characterization_2020}.
Work is also ongoing on alternative \ac{FET} configurations with lower intrinsic noise, allowing to achieve sub-electron readout noise levels with fewer skipper samples~\cite{sofo-haro_achieving_2024}.
Finally, the baseline for a future upgrade of the \ac{DESI} blue-channel CCDs are so called \ac{MAS} CCDs, where instead of using a single amplifier to measure a charge packet multiple times, multiple amplifiers are used to acquire multiple independent measurements of the same charge packet~\cite{lin_multi-amplifier_2024}.
This approach trades off the increase in readout time of classical Skipper CCDs with an increase in readout channels of the \ac{MAS} CCDs.
In this contribution we explore the relationship between the readout noise and readout time of classical Skipper CCDs without any hardware modifications, and we discuss the optimal trade-off between \ac{CDS} integration time and the number of skipper samples $N_{Skip}$ to achieve a given readout noise level with a minimal readout time.
The existence of such an optimal operating point has been demonstrated empirically in the context of the Astroskipper project~\cite{villalpando_characterization_2024}; here we expand on that work by discussing the trade-off from first principles, based on the intrinsic noise \ac{PSD}, which allows the optimum to be predicted rather than measured for each individual device and reveals that its location is governed by the per-sample charge-transfer time $t_{Skip}$.
This is complementary to other approaches for reducing the Skipper readout time, such as minimizing the per-sample dead time through optimized clocking sequences~\cite{lapi_fast_2022} and optimal digital filtering of the video signal~\cite{alessandri_optimal_2016}, as well as the \ac{ROI}, alternative-amplifier and \ac{MAS} approaches discussed above.

\section{Skipper CCD Readout Model}
\label{sec:model}
The readout of a (Skipper) CCD can broadly be divided into five different operations: parallel shift of $N_R$ rows, for each row $N_C$ horizontal shifts in the serial register, shifting each pixel charge into and out of the sense register $N_{Skip}$ times, integrating the baseline and charge for each pixel $N_{Skip}$ times, and the draining of each pixel charge packet.
The effect of adding horizontal or vertical overscan regions can be absorbed into $N_R$ and $N_C$.
Thus we can approximate the total readout time as
\begin{equation}
    t_{RO} =
        \underbrace{N_R \cdot t_{Vert}}_{\text{row shift}} +
        \underbrace{N_R \cdot N_C \cdot t_{Hor}}_{\text{serial shift}} +
        \underbrace{2 \cdot N_R \cdot N_C \cdot N_{Skip} \cdot t_{Skip}}_{\text{sense register shift}} +
        \underbrace{2 \cdot N_R \cdot N_C \cdot N_{Skip} \cdot t_i}_{\text{CDS integration}} +
        \underbrace{N_R \cdot N_C \cdot t_{Drain}}_{\text{drain}}
    \label{eq:readout_time}
\end{equation}
The row shift, serial shift and drain terms are not influenced by the use of skipping and have no direct influence on the CCD readout noise.
Thus any optimization of the readout time by using skipping, involves the sense register shift, and the \ac{CDS} integration time.

The readout noise as $\sigma_{RO}(t_i, N_{skip})$ can not be expressed as a simple closed form expression.
But it can be calculated numerically, by considering the CCD readout noise \ac{PSD} and by treating (Skipper) CCD readout via \ac{CDS} as a transfer function acting on the underlying noise \ac{PSD}.
\ac{CDS} effectively calculates the difference between baseline integration (i.e. the signal $s(t)$ at the readout node without any charge present) and signal integration (i.e. the signal $s(t)$ at the readout node with a constant charge present).
We assume that the signal can be decomposed into two uncorrelated components $s(t) = s_{charge}(t) + s_{noise}(t)$, with $s_{noise}(t)$ stationary.
Under this condition we can ignore $s_{charge}(t)$ for calculating the readout noise, and the application of \ac{CDS} to $s_{noise}(t)$ is a linear and time-invariant process.
Thus the effect of \ac{CDS} can be expressed as a transfer function in the frequency domain
\begin{align}
    h_{CDS}(f) &= -\frac{2j}{t_i} \frac{e^{-j \pi f t_i} e^{-j \pi f (t_i + t_{Skip})}}{\pi f} \sin(\pi t_i f) \cdot \sin(\pi (t_i + t_{Skip}) f) \\
    |h_{CDS}(f)| &= \frac{2}{\pi t_i f} \sin(\pi t_i f) \cdot \sin(\pi (t_i + t_{Skip}) f). \label{eq:cds_transfer_abs}
\end{align}
In literature the time needed to shift charges in and out of the readout node $t_{Skip}$ (often denoted as $t_{Clk}$) is often ignored when developing the \ac{CDS} and skipper transfer function~\cite{fernandez_moroni_sub-electron_2012,stefanov_optimal_2014,cancelo_deep_2011,cancelo_low_2020}.
This approximation is only valid if $t_{Skip} << t_i$, which is explicitly not the case for the readout of the Oscura sensor used in the next section.
With $t_{Skip} \approx t_i$ the true transfer function according to \eqref{eq:cds_transfer_abs} is noticeably shifted to lower frequencies, compared to the approximate calculation with $t_{Skip} = 0$.
Readout using skipper multi-sampling consists of multiple \ac{CDS} samples shifted in time-domain by $\Delta{}t = 2(t_i + t_{Skip})$\footnotemark, resulting in the frequency domain transfer function to be
\footnotetext{In the general case $\Delta{}t$ could take any value $\geq 2(t_i + t_{Skip})$.}
\begin{equation}
    h_{Skipper}(f) = \frac{1}{N_{Skip}} \sum_{i=0}^{N_{Skip} -1} h_{CDS}(f) \cdot e^{-j 2 \pi (i \Delta{}t) f}
\end{equation}
with no simple closed expression in the general case (i.e. with $t_{Skip} \neq 0$)~\cite{fernandez_moroni_sub-electron_2012}.

The readout noise of a CCD is dominated by the noise \ac{PSD} of the \ac{FET} used for sensing the charge packets.
Generally the \ac{PSD} of these devices is composed of two main components, white noise with a uniform spectral response and a $\nicefrac{1}{f}$ component which increases at lower frequencies, and can thus be modelled as
\begin{equation}
    S(f) = S_{white} \left(1 + \left(\frac{f_c}{f}\right)^\alpha \right), \label{eq:psd_model}
\end{equation}
with $S_{white}$ the white noise \ac{PSD} level, $\alpha$ an empirical exponential factor usually between \SIrange{1}{2}{}, and $f_{c}$ the frequency at which the white noise is equal to the $\nicefrac{1}{f}$ component~\cite{janesick_scientific_2001,fernandez_moroni_sub-electron_2012}.
The \ac{FET} of an \ac{FGA} is capacitively coupled to the CCD bulk and thus has a higher input capacitance compared to a \ac{FDA}, leading to a higher intrinsic noise \ac{PSD} of Skipper CCDs~\cite{fernandez_moroni_sub-electron_2012}.
Finally we can calculate the average readout noise based on the Wiener–Khinchin theorem as
\begin{equation}
    \sigma_{RO}(t_i, N_{Skip}) = \sqrt{\int_{0}^\infty \left|h_{Skipper}(f)\right|^2 S(f) df}, \label{eq:sigma_ro}
\end{equation}
treating $S(f)$ as a one-sided noise \ac{PSD}~\cite{janesick_scientific_2001,stefanov_optimal_2014}.
It should be noted, that the skipper transfer function acts as a low pass filter, thus ensuring the convergence of the integral.
In the following we evaluate these expressions numerically using the Python programming language.

\section{Application to an Oscura Microchip Skipper CCD}
\label{sec:measurement_oscura}
\begin{figure}
    \centering
    \begin{subfigure}[t]{0.49\textwidth}
        \includegraphics[width=\linewidth]{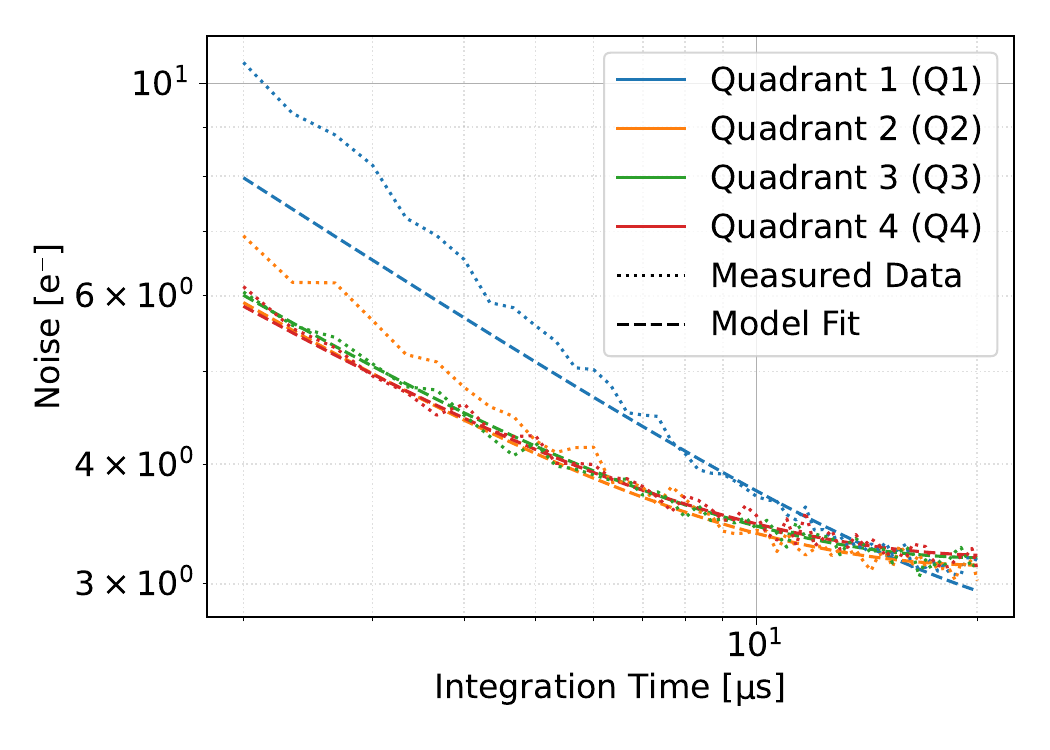}
        \caption{Measured and fitted $\sigma_{RO}(t_i)$ for $N_{Skip} = 1$.}
        \label{fig:psd_fitting_sigma}
    \end{subfigure}
    \hfill
    \begin{subfigure}[t]{0.49\textwidth}
        \includegraphics[width=\linewidth]{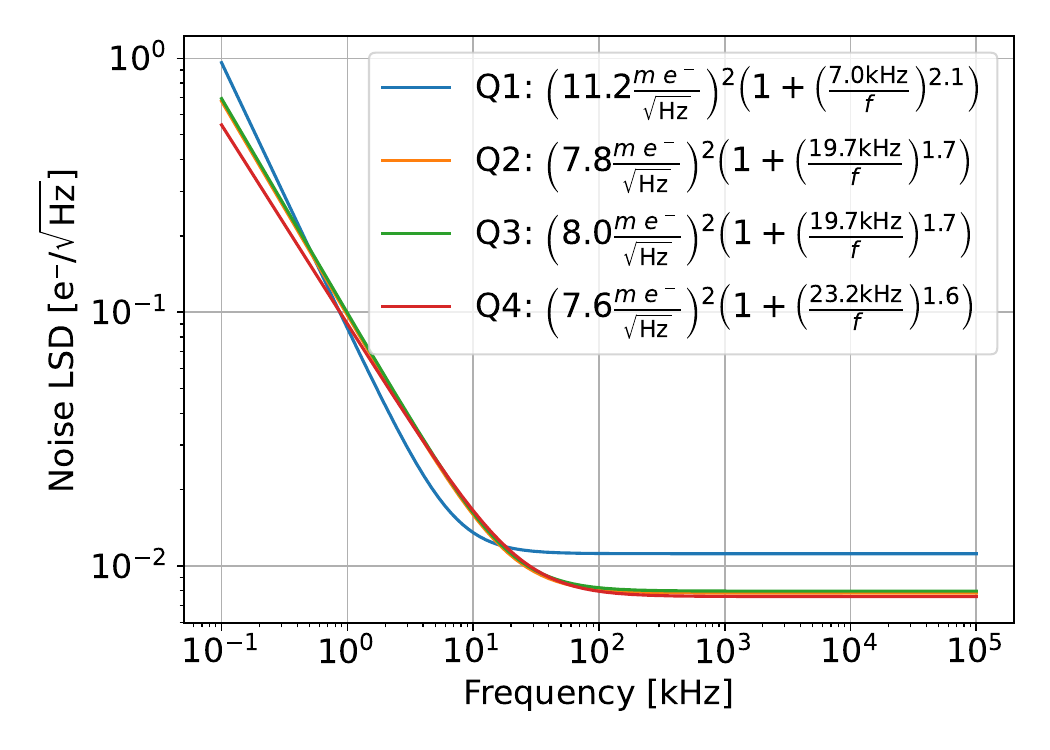}
        \caption{Extracted noise \ac{PSD} for each quadrant.}
        \label{fig:psd_fitting_psdextracted}
    \end{subfigure}
    \vspace{0.1cm}
    \caption{%
      Extraction of the underlying noise \ac{PSD} by fitting the noise model~\eqref{eq:sigma_ro} using data for $N_{Skip} \in [1, 2, 3, 5]$.
      Each CCD quadrant is fitted independently.
    }
    \label{fig:psd_fitting}
\end{figure}
\begin{figure}
    \centering
    \includegraphics[width=\linewidth]{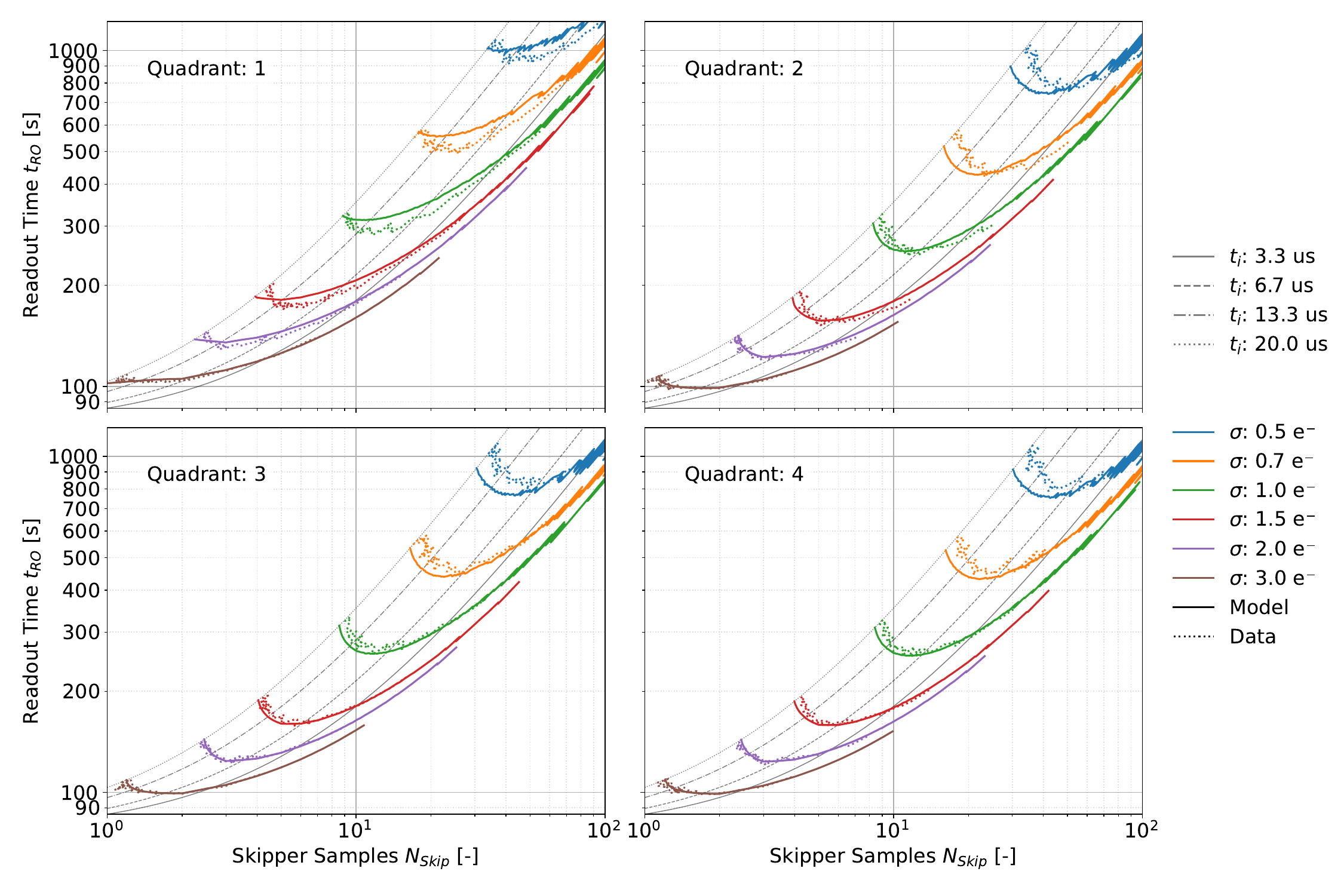}
    \caption{Simulated (solid lines) and measured (dotted lines) readout noise isocontours for the Oscura Microchip Skipper CCD.
    The data shows a clear minima in readout time for a given readout noise level.
    The isocontour lines for different values of $t_i$ are shown in grey.
    }
    \label{fig:oscura_isocontours}
\end{figure}

We apply the previously discussed read noise model to an Oscura-type Skipper CCD fabricated by Microchip~\cite{cervantes-vergara_skipper-ccd_2023}.
The sensor is made of four quadrants each with a dedicated readout channel, with overall \qtyproduct{1278x1058}{pixel}.
The \ac{DUT} is operated on a dedicated test stand, which operates the sensor at \SI{140}{\kelvin}.
The device is operated in dark conditions.%
\footnote{The test stand would allow to expose the \ac{DUT} using monochromatic light, but this capability was not used in this study, and a cover was secured to the cryostat window.}
Control and readout is carried out using a \ac{LTA} board which has a maximum clocking and sampling frequency of \SI{15}{\mega\hertz}~\cite{cancelo_low_2020}.
The gain of each quadrant was extracted from measurements with $N_{Skip} = 1000$ from the separation between individual electron peaks.
This allows to express the noise in the following in units of $e^{-}$.
The noise \ac{PSD} of each quadrant is extracted by measuring the readout noise as a function of the \ac{CDS} integration time $\hat{\sigma}_{RO}^{N_{Skip}}(t_i)$ for each channel.
The readout noise is measured from pixel values during a reverse overscan, where pixel charges are clocked away from the readout node thus guaranteeing that no excess charges generated from thermal effects or from clocking influence the measured distribution~\cite{cuevas-zepeda_automating_2025}.
The readout noise model~\eqref{eq:sigma_ro} is then fitted to the measured data to extract the noise \ac{PSD}~\eqref{eq:psd_model} for each quadrant independently.
For better numerical stability data with $N_{Skip} \in [1, 2, 3, 5]$ was used for the fitting.
Figure~\ref{fig:psd_fitting_sigma} shows the measured $\hat{\sigma}_{RO}^1(t_i)$ curves including the fitted noise model, and figure~\ref{fig:psd_fitting_psdextracted} the extracted noise \ac{PSD} for each quadrant.
From independent spectral measurements we know that the effective noise \ac{PSD} of the \ac{DUT} does not exactly follow the \ac{PSD} model in~\eqref{eq:psd_model}, mainly due to the presence of pickup noise at frequency \SI{>500}{\kilo\hertz}, which is due to non-ideal circuit and grounding design.
This excess noise in part explains the slight discrepancy between the measured and fitted curves in figure~\ref{fig:psd_fitting_sigma}, and is different for each of the four quadrants.
Quadrant 1 has the largest external noise pickup, and accordingly also the biggest discrepancy between the measured readout noise and the fitted noise model.
But as shown in the following, the simple noise \ac{PSD} model is sufficient to accurately predict the readout time as a function of the readout noise.
The readout time $t_{RO}(t_i, N_{Skip})$ is calculated using a clock accurate model mirroring the CCD clocking sequence employed.
Based on the extracted noise \ac{PSD} and \eqref{eq:sigma_ro} we calculate the readout noise for $\SI{670}{\nano\second} \leq t_i \leq \SI{20}{\micro\second}$ and $1 \leq N_{Skip} \leq 300$.
Calculating the corresponding readout times allows to plot readout noise isocontour lines in the $N_{Skip}$ vs. $t_{RO}$ plane.
The resulting readout noise curves are plotted in figure~\ref{fig:oscura_isocontours}.
For a given readout noise level $\sigma_{RO}$ a clear combination of $t_i$ and $N_{Skip}$ for which the readout time is minimal is visible.
For the \ac{DUT} used this minimum is close to $t_i \approx \SI{9}{\micro\second}$ with slight variations between the different quadrants.
It should be noted, that the value of $t_i$ at which the minimal readout time occurs is dependent on the skipper clocking time.
Further, at the optimal value of $t_i$, the noise has not yet converged to the minimal achievable noise without using skipping, as seen in figure~\ref{fig:psd_fitting_sigma}.
Both of these aspects are further discussed in the next section.
To validate the accuracy of the readout noise predicted by the model, the readout noise was directly measured with the Oscura Skipper CCD for $\SI{2}{\micro\second} \leq t_i \leq \SI{20}{\micro\second}$ and $1 \leq N_{Skip} \leq 100$.
The measured readout noise isocontour lines are also plotted in figure~\ref{fig:oscura_isocontours} as dotted lines.
A good agreement to within \SI{20}{\percent} of the readout time between the predicted and measured isocontour lines over the entire measured range can be seen, with the measured data also showing clear minima in the readout time for a given readout noise level.

\section{Discussion}
\label{sec:discussion}
The results from the previous sections imply that one can improve the overall readout time by operating the Skipper CCD with shorter \ac{CDS} integration times which lead to higher \textit{intrinsic} readout noise, and then recover the loss in readout noise via Skipper multi-sampling.
We can illustrate this effect with an intuitive example, by looking at the $\sigma^1_{RO} ~ \text{vs.} ~ t_i$ curves in figure~\ref{fig:psd_fitting_sigma}.
We observe for example for quadrant 3, that increasing the integration time by a factor 10 from \SI{2}{\micro\second} to \SI{20}{\micro\second} decreases the readout noise only by a factor of ca.~$2$.
Comparing this to the effect of increasing the number of Skipper samples by a factor of 10, leading to a reduction in readout noise of~$\sqrt{10}\approx 3$ shows the apparent benefit of using more skipper samples instead of longer integration times to reach a given readout noise level.
From a noise \ac{PSD} perspective this gain comes from the fact, that increasing $t_i$ will shift main lobe of the transfer function towards lower frequencies, and thus in~\eqref{eq:sigma_ro} the $\nicefrac{1}{f}$ component of the noise \ac{PSD} is more dominant.
On the other hand, increasing $N_{Skip}$ does not shift the position of the transfer function main lobe, rather decreases the width of the lobe, and thus does not lead to an increase in integrated $\nicefrac{1}{f}$ noise.
\begin{figure}
    \centering
    \begin{subfigure}[t]{0.49\textwidth}
        \centering
        \includegraphics[width=0.95\linewidth]{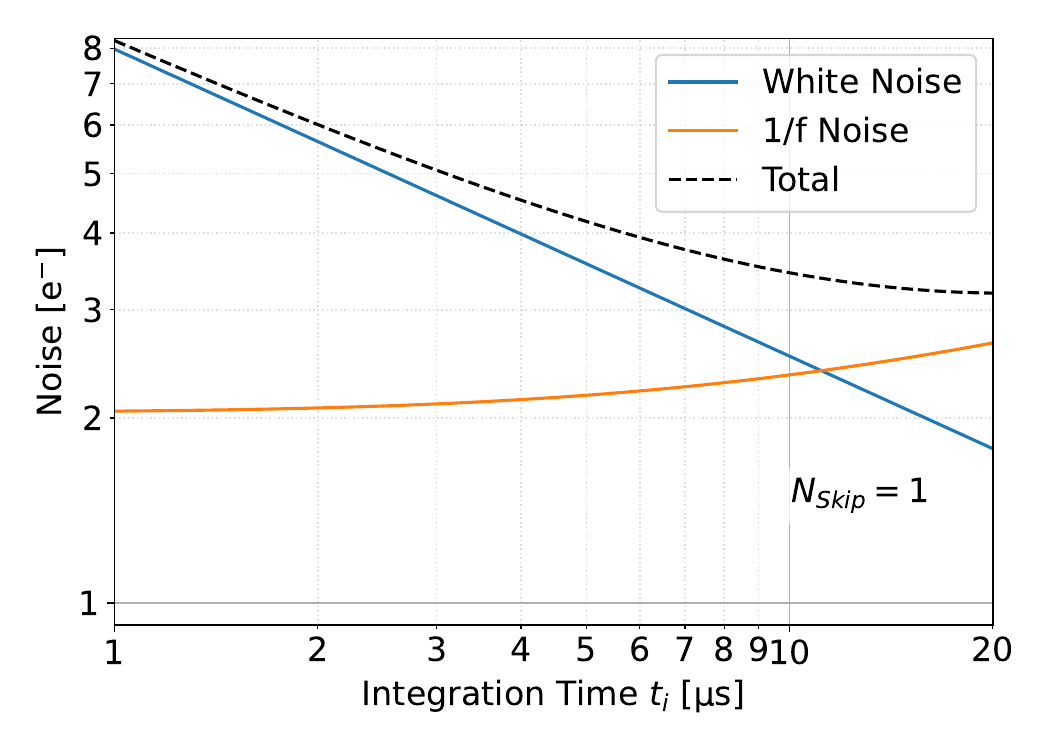}
        \caption{Readout noise components for~$N_{Skip} = 1$ in function of the \ac{CDS} integration time~$t_i$.}
        \label{fig:noise_components_ti}
    \end{subfigure}
    \hfill
    \begin{subfigure}[t]{0.49\textwidth}
        \centering
        \includegraphics[width=0.95\linewidth]{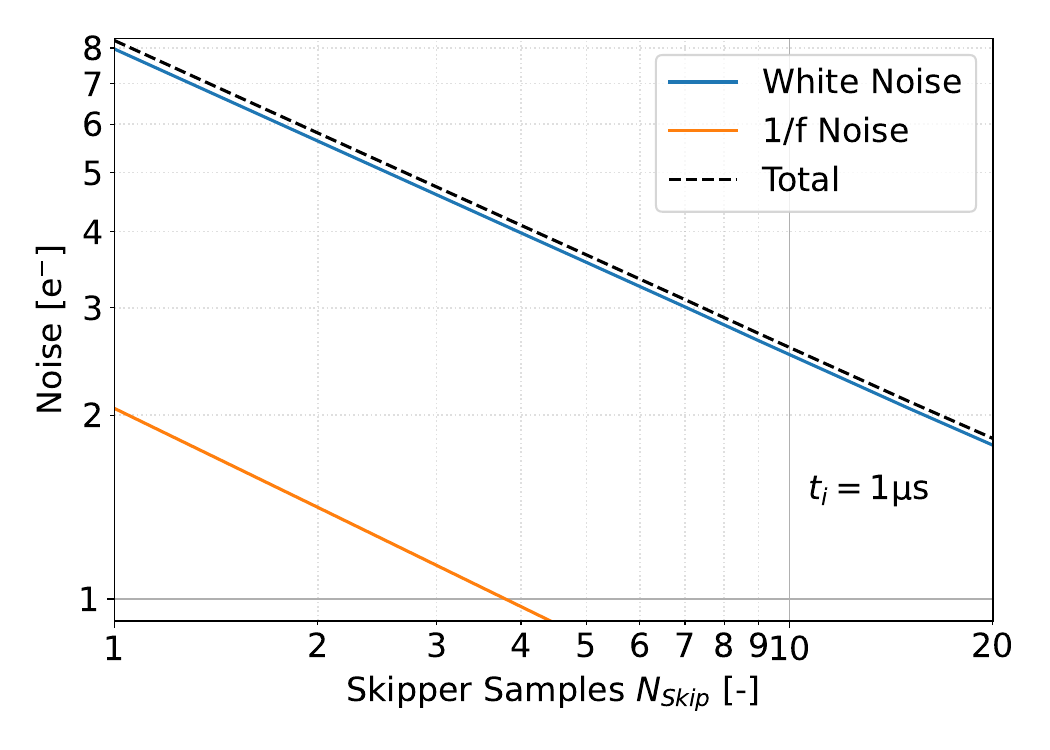}
        \caption{Readout noise components for~$t_i = \SI{1}{\micro\second}$ in function of~$N_{Skip}$.}
        \label{fig:noise_components_nskip}
    \end{subfigure}
    \vspace{0.1cm}
    \caption{Individual readout noise components (white noise and $\nicefrac{1}{f}$ noise) as predicted based on \eqref{eq:sigma_ro}, using the extracted noise \ac{PSD} of quadrant 3 of the Oscura Skipper CCD.
    The parameters ($N_{Skip} = 1$ in figure~(a) and $t_i = \SI{1}{\micro\second}$ in figure~(b)) are chosen such that the total integration time $N_{Skip} \cdot t_i$ is equal for the respective points in the two plots.
    One can see that the noise from the white noise component only depends on this total integration time, and not on whether skipper readout is used or not, the same is not true for the $\nicefrac{1}{f}$ component.
    Note: The individual noise components add in quadrature to the total noise.
    }
    \label{fig:noise_components}
\end{figure}
This is illustrated in figures~\ref{fig:noise_components_ti}, where increasing $t_i$ leads to a larger $\nicefrac{1}{f}$ component, which eventually dominates the readout noise.
On the other hand increasing $N_{Skip}$, as shown in figure~\ref{fig:noise_components_nskip} does not lead to a relative increase of the $\nicefrac{1}{f}$ component.
Thus one is compelled, to prioritize an increase $N_{Skip}$ with smaller $t_i$ to improve the readout noise.
But from the readout time perspective, this only holds as long as the readout time is dominated by the integration time $t_i$ and not by the time needed to shift charges in and out of the sense register $t_{Skip}$.
Once $t_i$ becomes comparable to $t_{Skip}$ the benefit of adding further skipper samples starts to diminish, as the time cost of each additional sample becomes dominated by the fixed $t_{Skip}$ overhead rather than by $t_i$, and eventually the readout time starts to increase again.
The optimal integration time is therefore set by the scale of the charge-transfer time $t_{Skip}$ (together with the shape of the noise \ac{PSD}).
Since both noise components scale as $\nicefrac{1}{\sqrt{N_{Skip}}}$, the target noise level sets only $N_{Skip}$ and not the location of this optimum.
The optimal $t_i$ is therefore essentially independent of the target readout noise, as observed in figure~\ref{fig:oscura_isocontours}.
\begin{figure}
    \centering
    \includegraphics[width=\linewidth]{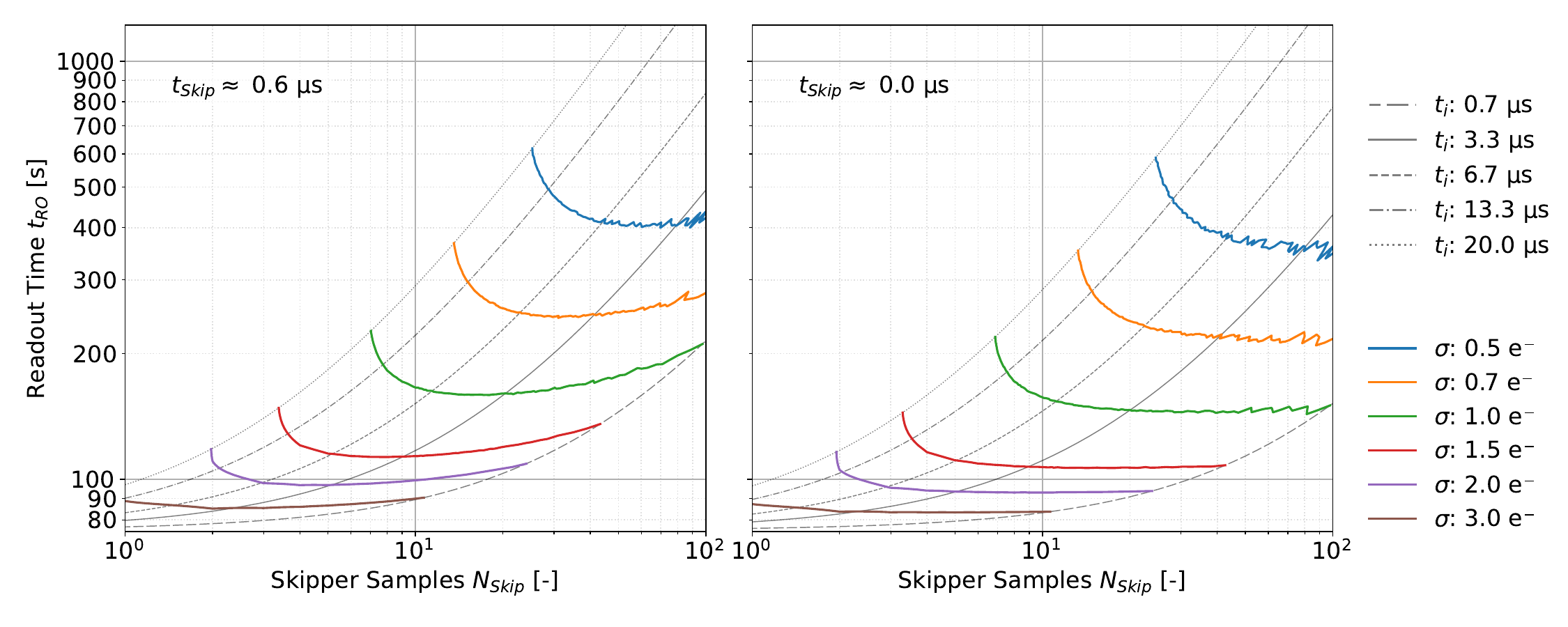}
    \caption{Simulated readout time and readout noise for two cases with shorter $t_{Skip}$.
    Reducing $t_{Skip}$ leads to a shift of the optimal readout time towards lower integration times $t_i$.
    In the extreme case of $t_{Skip} \rightarrow \SI{0}{\micro\second}$ the minimum disappears entirely: decreasing $t_i$ while increasing $N_{Skip}$ then reduces the readout time asymptotically, with no optimal $t_i$.
    }
    \label{fig:oscura_extrapolated}
\end{figure}
To showcase this effect, we simulated the readout noise and readout time for quadrant 3 of the Oscura sensor with an assumed much shorter $t_{Skip} \approx \SI{0.6}{\micro\second} ~ \text{and} ~ \SI{0.0}{\micro\second}$.
The resulting readout noise isocontour lines are shown in figure~\ref{fig:oscura_extrapolated}.
As expected the overall readout time is lower with shorter skipping transfer times $t_{Skip}$.
But, as shown in figure~\ref{fig:oscura_extrapolated}, shorter transfer times also shift the minimum of the isocontour curves towards shorter \ac{CDS} integration times, and thus higher values of $N_{Skip}$ for the optimal readout time at a given noise level.
In the extreme case of $t_{Skip} \approx \SI{0.0}{\micro\second}$ there is no longer an optimal $t_i$ for which the readout time is minimal, rather, decreasing $t_i$ while increasing $N_{Skip}$ asymptotically decreases the readout time.
This is consistent with the previous considerations, as in this case there the readout time can be kept constant with $t_i \propto \nicefrac{1}{N_{Skip}}$.

These results imply that the optimization and reduction of the $t_{Skip}$ is absolutely essential to enable fast Skipper CCD readout.
Evidently, in a real system the $t_{Skip}$ time can not be arbitrarily low.
From a technical perspective the time $t_{Skip}$ includes the transfer of charges in or out of the sense node, the reset of the \ac{FGA} input, but also the wait time necessary for the output to settle into a steady state.
All of these operations have absolute minimal execution times bounded by physical constraints.
In addition, certain operations are usually artificially slowed down to reduce secondary effects.
The \ac{LTA} board for example is equipped with low pass filters to increase to control $\nicefrac{dV}{dt}$ and cross-talk~\cite{cancelo_low_2020}.
In addition the board is built around a \SI{15}{\mega\hertz} sampling frequency and thus has appropriate anti-aliasing filters on the video lines, leading to relatively long settling times for the output signals.
It should be noted that the \ac{LTA} board was not built with fast Skipper readout in mind, but rather to achieve the absolutely lowest readout noise levels possible, as desired for dark matter searches~\cite{cancelo_low_2020}.
In order to truly establish Skipper CCDs as tool for astronomy, dedicated optimized Skipper CCD readout systems are needed.
These systems need to enable fast CCD clocking times, as well as very short \ac{CDS} integration times by providing high enough bandwidths of the video channels.
More specifically, fast skipper clocking is essential, the clocking speed of vertical and horizontal transfers are secondary.
Such fast skipper clocking speeds can of course lead to detrimental secondary effects, such as \ac{CIC}, \ac{CTI}, cross-talk and amplifier glow, which need to be studied in detail for fast operating output stages.

\section{Conclusions}
\label{sec:conclusions}
The non-destructive multi-sampling of Skipper CCDs allows reducing readout noise to sub-e$^-$ levels, making Skipper CCDs single-electron and single-photon sensitive.
For astronomical applications this noise reduction is highly attractive for observations of faint sources, but unlike in dark matter searches the associated increase in readout time cannot be ignored, as it directly competes with the available exposure time.
The central trade-off is therefore between the \ac{CDS} integration time $t_i$ and the number of Skipper samples $N_{Skip}$ used to reach a target readout noise level.

We have modelled this trade-off from first principles by treating Skipper readout as a transfer function acting on the intrinsic noise \ac{PSD} of the sensor.
This model shows that, because increasing $N_{Skip}$ suppresses the readout noise without shifting additional $\nicefrac{1}{f}$ noise into the signal band, multi-sampling at short integration times is generally favourable over longer integration times.
As a consequence, for any given target readout noise there exists a combination of $t_i$ and $N_{Skip}$ that minimizes the total readout time.
The existence of such an optimum has previously been observed empirically~\cite{villalpando_characterization_2024}.
The present work explains its origin and, crucially, allows the optimal operating point to be predicted from the measured noise \ac{PSD} of a device rather than determined experimentally for each CCD architecture.
We validated this prediction on an Oscura Microchip Skipper CCD, finding good agreement between the modelled and measured readout-time isocontours over the full range of operating parameters.

Our analysis further reveals that the location of this optimum is governed by the per-sample charge-transfer time $t_{Skip}$: the benefit of additional Skipper samples diminishes once $t_i$ becomes comparable to $t_{Skip}$, and in the limit $t_{Skip} \rightarrow 0$ the optimum disappears entirely, with ever shorter $t_i$ and larger $N_{Skip}$ asymptotically reducing the readout time.
Minimizing $t_{Skip}$ is therefore key to fast Skipper CCD readout and motivates dedicated readout electronics optimized for fast skipper clocking and short \ac{CDS} integration times, rather than for the lowest achievable noise as in current dark-matter-oriented systems.

\appendix

\acknowledgments
We thank Juan Estrada, Alex Drlica-Wagner, Brenda Cervantes Vergara, Claudio Chavez and Brandon Roach from Fermilab for providing us the Oscura Skipper CCD, and for valuable inputs on operating scientific CCDs.
This work is supported by the University of Zurich.

\bibliography{20260529_SPIEPaper} 
\bibliographystyle{spieref} 

\end{document}